\documentclass[10pt]{iopart}

\usepackage{times}
\usepackage{graphicx}
\usepackage{amsfonts}
\usepackage{amssymb}
\usepackage{hyperref}
\usepackage{wasysym}
\usepackage{flafter}

\begin{document}
\title{Pressure sensing by electro-mechanical coupling in compliant dielectric membranes polarized by a bias voltage}

\author{B Van Damme$^{1}$, A Brun d'Arre$^{1}$, P Danner$^2$, D Opris$^2$, A Bergamini$^{1}$}

\address{$^{1}$ Empa-Materials Science and Technology, Acoustics/Noise Control\\
Ueberlandstrasse 129, CH-8600 Duebendorf, Switzerland}
\address{$^{2}$ Empa-Materials Science and Technology, Functional Polymers\\
Ueberlandstrasse 129, CH-8600 Duebendorf, Switzerland}
\ead{bart.vandamme@empa.ch}	

% \keywords{Electromechanical coupling, Waves, Soft Actuators, Piezoelectric, Polarization}

\begin{abstract}
Among smart materials, piezoelectric materials occupy a very prominent position for sensing and actuation functions. Combined with simple or more advanced shunts, they are also proposed in various vibration mitigation schemes. However, the selection of available piezoelectric materials is mainly limited to ceramics (with an elastic modulus in the order of $10^{10}Pa$ (e.g. PZT ceramics) and a few polymer materials, with elastic modulus in the range of $10^{9}Pa$ (e.g. PVDF). In both cases, the high mechanical impedance and, consequently, the small dynamic strains limit the application of these materials to stiff structures. In this contribution, we discuss using a bias voltage to polarize dielectric materials and thereby compensate for the lack of spontaneous polarization observed in piezoelectrics. This enables access to materials with a wider range of elastic properties, such as soft elastomers, e.g. poly(dimethylsiloxane). As an example, we present a practical implementation of a silicone rubber membrane used as a highly compliant dynamic pressure sensor. For such nearly-incompressible materials, the capacitance change during dynamic deformation of the membranes is sufficiently large to generate a measurable dynamic voltage change over the membrane.
\end{abstract}

\submitto{\SMS}
\maketitle
\ioptwocol
% \maketitle

\section{Introduction}
Energy coupling between the mechanical and electrical domain using smart materials offers a host of possible applications, ranging from strain and vibration sensing to shape morphing and energy harvesting~\cite{bahl2020smart,khoo20153d,safaei2019review}. Piezoelectric materials hold a prominent position in this field, due to their excellent coupling coefficients (e.g. lead zirconate titanate ceramics or PZT) and are widely used in applications ranging from consumer goods to precision instruments~\cite{sekhar2023review,habib2022review}. Alternatively, the change of capacitance in compliant mechanical systems can be used to detect external stimuli, as is the case in membrane microphones (for pressure sensing) and MEMS sensors (for pressure or acceleration)~\cite{cheng2023recent,qin2021flexible}. A mechanical excitation changes the distance between electrodes separated by a dielectric medium, typically air, which leads to a flow of charges through a shunt resistor. All these systems have in common that the output voltage, either over the piezo or over the shunt resistor, is proportional to the measured quantity, which allows accurate calibration over relatively wide frequency ranges. 

The material properties of ceramic piezoelectrics make them ideal candidates for applications in stiff structures with a high mechanical impedance, where the strains are compatible with the modest strain capacity of the materials ($\tau_{failure}\approx 1\permil$ \cite{munz1998deformation}). However, in some cases, small impedance values and large strains are desirable. For example, coupling acoustic power into electrical circuits or providing frequency dependent stiffness capabilities in the sense of what we presented in recent work \cite{van2024implementation} to systems subject to large strains, would benefit from electromechanical coupling elements with substantially lower elastic modulus $E$ than ceramics ($E_Y\approx70$~GPa \cite{bouzid2005pzt}) or engineering polymers, such as PVDF ($E_Y\approx2.5$~GPa \cite{dmitriev2006dependence}). Everyday engineering systems such as engine suspensions or shock dampers undergo much larger strains, and the growing field of soft robotics has a high demand for compliant materials. 

\begin{figure}
     \centering
     \includegraphics{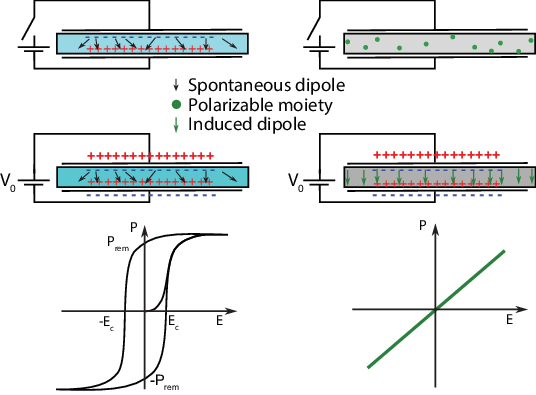}
     \caption{Piezo-electric materials (left) are characterized by the presence of spontaneous dipoles that are oriented during the polarization process. In linear dielectrics (right), dipoles are induced by external fields but are not present in absence of electric fields.}
     \label{fig:Micro_Macro-dipole}
 \end{figure}

While recent efforts \cite{owusu2023make} have led to the development of piezoelectric elastomers, in this paper we present an alternative to such highly specialized materials. An external bias voltage applied over elastomer membranes can achieve macroscopic polarization and realize strong electromechanical coupling. This approach warrants great flexibility in selecting dielectric materials, based on their elastic properties, and allows for tuning the polarization and electromechanical coupling, via the applied bias voltage. Fig.~\ref{fig:Micro_Macro-dipole} shows a conceptual comparison between the spontaneous polarization of ferroelectrics and the polarization from an applied bias voltage.

In piezoelectric materials (polarization curve shown in the bottom left of Fig.~\ref{fig:Micro_Macro-dipole}), the spontaneous dipoles observed at $E=0$~V/m are structurally coupled to the strain of the material, via the distance between the centers of charge in the crystal (such as in the PZT Perovskite structure), making the polarization strain dependent. In a linear dielectric however, no spontaneous polarization is observed. Dipoles can be induced by applying an external field $E$, as shown in Fig.~\ref{fig:Micro_Macro-dipole}. In the presence of a bias electric field, the change in thickness because of the deformation of a compliant dielectric will cause a change of the polarization of the system associated with the change of its capacitance. Consequently, as a dynamic load deforms the dielectric, a current will flow to the electrodes to balance the charges to satisfy $Q(t)=C(t)/V_{bias}$, where $Q(t)$ is the charge accumulated on the electrodes, $C(t)$ is the time variable capacitance of the device, and $V_{bias}$ is the applied bias voltage, see eq.~(\ref{eq:current}).

Compliant devices for pressure sensing are extensively investigated, especially in the context of wearable systems such as e-skins \cite{Ha2022}, often for health monitoring \cite{Kim2019,} applications. The devices can be classified, based on their mode of operation, into piezo-resistive\cite{Chen2020}, capacitive, and piezo-electric\cite{Chen2020, Dagdeviren2014}. The former exploits changes in resistivity caused by the deformation of a compliant element. These devices can rely on the specific surface microstructure \cite{Zammali2022} of a compliant resistive element that changes specific resistivity upon deformation, for example due to percolation of conductive micro- or nano-particles \cite{Zhao2023}. Piezoelectric sensing relies either on specific elastomeric materials that contain dipoles and exhibit the necessary polarization \cite{Owusu2022} or on some kind of compounding that can happen at different length scale  \cite{jin2021skin,Dagdeviren2014}. Finally, capacitive sensing is based on a change of the dimensions of a capacitor device, as is the case for membrane microphones \cite{Kumar2024}, which can be realized at different length scales and exploiting different dielectrics ranging from air \cite{Shin2017} to solid \cite{Kumaresan2022, Zammali2022}.

Here, we describe a simple sensing device that exploits changes in capacitance but overcomes the limitation given by the lack of spontaneous polarization in conventional dielectric elastomer actuators. We describe models (Sec. 2) of and experiments (Sec. 3) on thin, compliant dielectric membranes, subject to a static bias voltage, and their coupling with an acoustic pressure field. The system discussed here shows that strong electromechanical coupling can be achieved using standard silicone rubber with moderate relative permittivity. Such a membrane is the most basic component of potential sensing or smart devices that can be assembled from it. Arrays of membrane elements allow to couple acoustic energy into the electrical domain exploiting the equivalent of the '31' coupling in piezoelectrics over larger areas, while stacked compliant membranes are customarily used as stacked transducers, exploiting '33' coupling\cite{kovacs2009stacked}. 

\section {Modelling electro-mechanical coupling in dielectric membranes}
%We used numerical models (implemented in ANSYS) of a dielectric membrane to assess their dynamic response. The dimensions of the membranes investigated in the modelling effort were a radius of 25mm and a thickness of 0.1mm ($r_0$ and $d_0$ in the bottom left of fig. \ref{fig:Membrane_Model}). These dimensions correspond to the standard size of the samples that can be prepared in our laboratories, for materials testing purposes. A linear elastic modulus of $E_Y \approx 1.7$~MPa and a Poisson's ratio of 0.49 were assumed for the models, corresponding to a generic Polydymethylsiloxane. The pre-stretching of the membrane was accounted for by carrying out a static pre-stress step, with an imposed radial deformation of 5~mm, that was used as starting condition for the subsequent modal analysis and associated harmonic response, assuming a nodal pressure applied to the surface of the membrane. Indications on the variation of the membrane capacitance in relation to the resonant deformations of the membrane were an additional output of the numerical investigation. Here, a simplification was made to estimate the area and thickness change of the membrane by approximating the curved membrane with a conical surface, described in the bottom right of fig.\ref{fig:Membrane_Model}. 
\subsection{Analytical approximation of dynamic voltage generation}
In this work, a nearly incompressible membrane is first stretched to a desired mechanical prestrain, and then polarized by a DC bias voltage charging the electrodes on both sides of the membrane. The electrostatic force acts on the electrodes on either side of the membrane, thereby slightly relaxing the static mechanical strain but still resulting in a steady capacitance of the membrane defined by the membrane's area $S=\pi r_0^2$ and thickness $d_0$ between the electrodes. When a dynamic force or pressure is applied to the membrane's surface, the out-of-plane deformation of the membrane leads to an important increase of the area. Assuming a perfectly incompressible material, this results in a reduced thickness and, hence, an increased capacitance.

\begin{figure}
     \centering
     \includegraphics{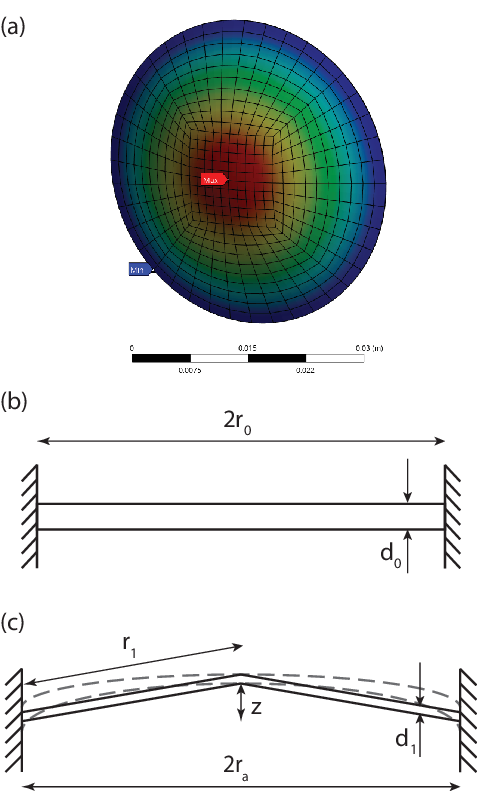}
     \caption{ANSYS (2023 R1) simulation of the shape corresponding to the 1st mode of the membrane at a frequency of approximately 308~Hz (a). The change of surface area and thickness from undeformed (b) to deformed (c), based on the assumption of volume invariance, leads to a variation of the membrane's capacitance. }
     \label{fig:Membrane_Model}
 \end{figure}
 
 Fig.~\ref{fig:Membrane_Model}(a) shows the  membrane mode targeted in this work, aiming at coupling acoustic power into the the electrical domain. The  change of capacitance of the membrane caused by its deformation can be deducted from simplified geometric considerations, namely as a function of the out of plane deflection $z$. The modal shapes of the membrane were computed using ANSYS 2025R2. The membrane was segmented in a central panel (square) and four segments around the circumference to obtain quadrilateral elements with small aspect ratio. The indices in Fig.~\ref{fig:Membrane_Model}(b) and (c) refer to the following membrane states:
\begin{itemize}
\setlength\itemsep{0em}
    \item index \(0\): undeformed, non-strained membrane, with known radius \(r_0\) and thickness \(d_0\),
    \item index \(a\): strained membrane, with known radius \(r_a\),
    \item index \(1\): deformed, strained and clamped membrane, with known deflection \(z(t)\).
\end{itemize}

Straining the membrane from radius $r_0$ to $r_a$ reduces the tickness from $d_0$ do $d_a  =r_0^2 d_0 / r_a^2$ to conserve the volume. This will increase the membrane's capacitance from $C_0 = \frac{\varepsilon S_0}{d_0}$ to $C_a = \frac{\varepsilon S_a}{d_a}$.
The deformed shape of the membrane due to a perpendicular point force acting on its center is approximated as a cone with height $z(t)$, in which case the total surface equals $S_1 = \pi r_1 r_a$, where $r_1^2 = z^2 + r_a^2$. The final thickness of the membrane can be calculated from conservation of volume, and is given by $d_1 = r_0^2 d_0/(r_a r_1)$. Based on these geometrical relations, the final capacitance is given by
\begin{equation}
    C_1 = C_0 \frac{r_a^2}{r_0^4}\left( z(t)^2 + r_a^2\right),
\end{equation}
and the current flowing between the electrodes to account for the variation of the capacitance is:
\begin{equation}\label{eq:current}
    i_1(t) = V_0 \frac{dC_1}{dt} = 2V_0 C_0\frac{r_a^2}{r_0^4}z(t)\frac{dz(t)}{dt}.
\end{equation}
Here, $C_0$ is the capacitance of the undeformed membrane, $V_0$ is the applied bias voltage, and $z(t)$ is the out-of-plane displacement of the center of the membrane. In correspondence with the first vibration mode of the membrane, large displacement amplitudes of the center are expected to cause large capacitance variations.  

Assuming a harmonic motion $z(t) = z_0 \sin(\omega t)$, the current is given by
\begin{equation}\label{eq:current_harmonic}
    i_1(t) = 2V_0 C_0\frac{r_a^2}{r_0^4}z_0^2 \omega \sin(\omega t) \cos(\omega t) = V_0 C_0 \frac{r_a^2}{r_0^2} \frac{z_0^2}{r_0^2} \omega \sin(2 \omega t). 
\end{equation}
The resulting eq.~(\ref{eq:current_harmonic}) shows that the resulting current depends on the initial capacitance of the unstretched membrane, which combines the dielectric properties and thickness of the material, the static in-plane strain $\tau = r_a/r_0 - 1$, and the vibrational amplitude of the center of the membrane, divided by the membrane's radius. The generated current is proportional to the static bias voltage. Interestingly, the induced current fluctuates at double the frequency as the membrane's displacement. This is intuitively clear, since during one period of the displacement, the membrane will reach its minimal thickness twice. This is a fundamental difference from conventional capacitive sensors where the flexible electrode moves relatively to a rigid electrode, so that there is no frequency change between the driving mechanism and the measured current or voltage.

\subsection{Finite element analysis}
Since the analytical derivation in the previous section relies on some strong assumptions (perfectly incompressible materials, constant thickness of the membrane, approximation of the deformed shape), we model the expected current via a more accurate finite element model. However, the frequency-upconversion of the electric current vs. membrane displacement leads to a fundamental nonlinear response which cannot be captured by conventional linear harmonic response simulations. Therefore, a transient time-domain simulation is implemented in ANSYS 2025R2. The material properties represent a typical silicone rubber, with Young's modulus $E = 1$~MPa, Poisson ratio $\nu = 0.495$, and density $\rho = 1000$~kg/m$^3$. The relative permittivity is $3$. Since dielectric materials are not available within the Ansys framework, the membrane is defined as a piezoelectric material with all coupling values in the piezoelectric coupling matrix values set to 0. The initial thickness of the membrane is $d_0 = 0.1$~mm, and the initial radius is $r_0 = 20$~mm.

Although membrane dynamics are typically investigated using shell elements to avoid shear locking, this physical problem requires separated nodes on the top and bottom of the membrane to allow for a different voltage and charge separation. Therefore, a 3D geometry and mesh have to be defined for the Coupled Field Transient simulation. The top side of the membrane is defined as a full-surface electrode by coupling the voltage degree of freedom, the bottom side is a full-surface electrode connected to the ground through a resistor $R$ represented by a single CIRCU94 element. 

The transient simulation is defined through 3 loading steps. In step 1, the membrane is prestretched by a uniform radial displacement which for this mesh leads to converged solutions up to 5~mm deformation, representing a strain of 25\%. During step 2, the bias voltage is applied on the top electrode, and the membrane can be given an initial out-of-plane deformation by a static force applied over the membrane's surface. The duration of both steps is set to 1~s, with a time step of 5~ms. The dynamic response to a time-varying pressure load is simulated by a short triangular pressure pulse with amplitude 100~Pa, applied to the entire surface of the membrane. The duration of the pulse defines the maximum frequency and is set to 1~ms. The response is calculated over a 1.5~s time span, with a maximum time step of 0.1~ms -- ensuring at least 10 time steps within the shortest period that can be generated by the excitation pulse. Alternatively, the pulse can be replaced by a sweep or white noise pressure signal, but this requires typically longer simulated times or more time steps.

This numerical setup allows for a parametric investigation of five relevant quantities: the static prestrain $\tau_0$, the bias voltage $V_0$, the static out-of-plane membrane force $F_0$, the electrical resistance $R$, and the material's relative permittivity $\epsilon_r$. The out-of-plane force is introduced in order to investigate the voltage generation in a membrane with a structural imperfection that induces a deflection from the neutral plane, which is inevitably the case for the following experimental study. The chosen nominal values are $\tau_0 = 0.25$, $V_0 = 500$~V, $F_0 = 1$~mN, $R = 100~\Omega$, and $\epsilon_r = 3$. For each simulation, the time series of the displacement at the center of the membrane, and the voltage over the resistor are saved.

\begin{figure}
    \centering
    \includegraphics[width=\columnwidth]{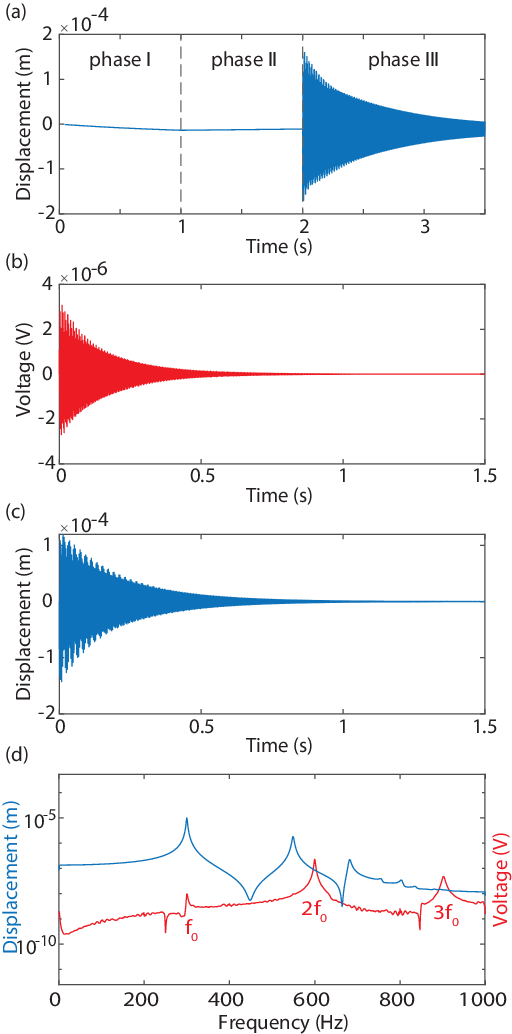}
    \caption{(a) Time signal over the three steps of the FE simulation: (I) static stretching, (II) out-of-plane loading and bias voltage, and (III) dynamic deformation due to a pressure pulse. (b)-(c) Windowed response of the dynamic displacement and voltage time signals, used for the calculation of the spectrum. (d) Spectra of membrane displacement and generated voltage, showing that the principal voltage frequency component is twice the fundamental displacement mode frequency.}
    \label{fig:timesig}
\end{figure}

A typical response of the voltage and displacement in the time domain is shown in Fig.~\ref{fig:timesig}, showing the three phases of the simulation. The prestrain results in a small reduction of the membrane's thickness, after which the entire membrane is deformed statically by the out-of-plane force. The pressure pulse then makes the membrane vibrate, leading to a decaying response over time. The power spectrum of these signals is calculated in Matlab using the built-in Welch' algorithm (the periodogram function), using an additional exponential window to reduce the signal's amplitude at the end of the simulation time, a usual procedure for impactive excitation. The windowed dynamic displacement and voltage signals are shown in Fig.~\ref{fig:timesig} (d), together with the displacement spectrum. The pressure pulse excites the first three modes of the membrane, all below 1000~Hz.

\begin{figure}
    \centering
    \includegraphics[width=\columnwidth]{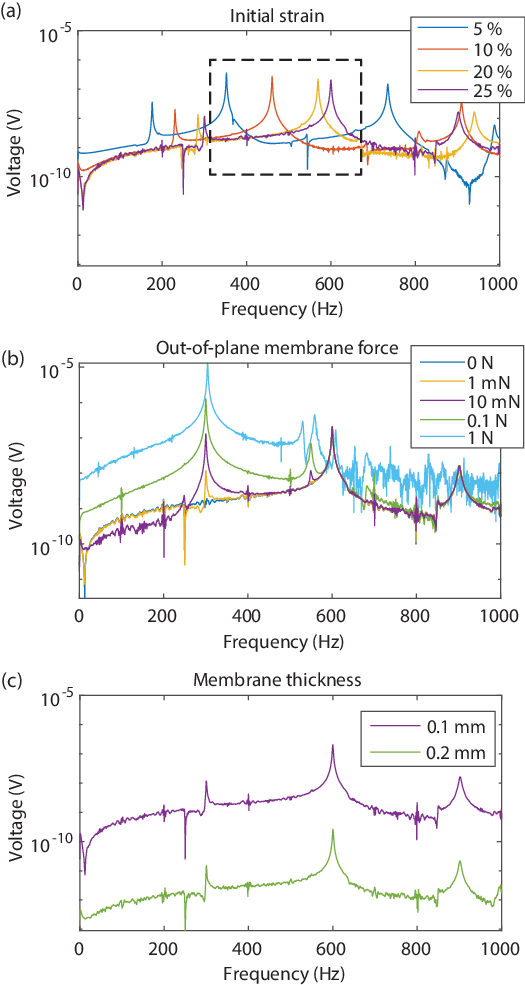}
    \caption{Voltage amplitude spectra for varying mechanical conditions of the membrane: initial in-plane strain, static out-of-plane membrane force, and membrane thickness.}
    \label{fig:mechanical-model}
\end{figure}

Three structural parameters are investigated, namely the in-plane strain, static out-of-plane membrane deformation from different force values, and thickness of the membrane. The results are shown in Fig.~\ref{fig:mechanical-model}. In case of increasing membrane strain, the eigenfrequencies of the membrane shift to higher frequencies. The peaks in the dashed frame in panel (a) are the highest-amplitude peaks, corresponding to the double frequency of the first membrane mode. The higher-frequency peaks are either higher harmonics of the first vibration mode, or correspond to higher deformation modes. Due to the initial out-of-plane deformation, the fundamental mode frequency is also represented in the voltage spectrum. This is clearly demonstrated in panel (b) where, for increasing out-of-plane static deformation, the voltage at the fundamental mode frequency increases strongly. At an initial membrane force of 1~N, which yields an out-of-plane deformation of 2.72~mm, the fundamental voltage peak is 30 times higher than the first harmonic. Finally, the spectra presented in panel (c) show that the membrane thickness has no influence on its eigenfrequencies. However, the higher mass and stiffness lead to lower displacement amplitudes, and the higher thickness to a lower initial capacitance $C_0$, which, combined, result in a much lower dynamic voltage generation. 

\begin{figure}
    \centering
    \includegraphics[width=\columnwidth]{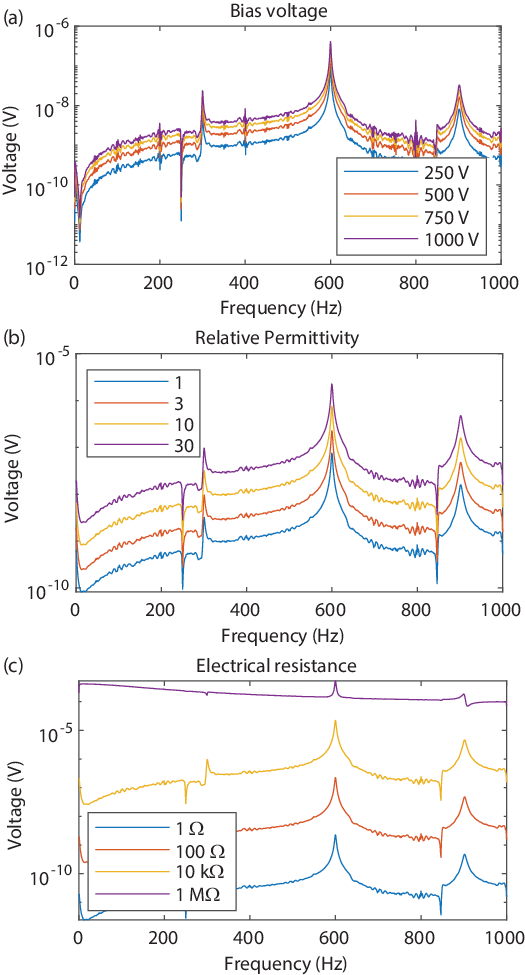}
    \caption{Voltage amplitude spectra for varying electrical conditions of the membrane: bias voltage, relative permittivity, and shunt resistance.}
    \label{fig:electrical-model}
\end{figure}

The electrical parameters under consideration are the bias voltage $V_0$, the membrane's permittivity $\varepsilon$, and the shunt resistor $R$, and the results are shown in Fig.~\ref{fig:electrical-model}. As predicted by eq.~(\ref{eq:current_harmonic}), the output voltage is proportional to the bias voltage and relative permittivity. Although the relative permittivity of silicone rubber membranes is typically relatively low, it is possible to manufacture dielectric membranes with a much higher permittivity~\cite{vonSzczepanski2023high}. Finally, the voltage over the shunt resistor is governed by the time constant $t_{RC} = RC$. As long as $t_{RC}$ is lower than the membrane's vibration period, the generated voltage increases proportionally to the resistance. However, if $t_RC$ is large, the low-pass filter properties of the RC-series shunt become visible, and the signal-to-noise ratio of the generated voltage peaks becomes significantly worse.

\section{Experimental investigation}
\subsection{Sample Preparation}
\begin{figure*}
     \centering
     \includegraphics[width = \textwidth]{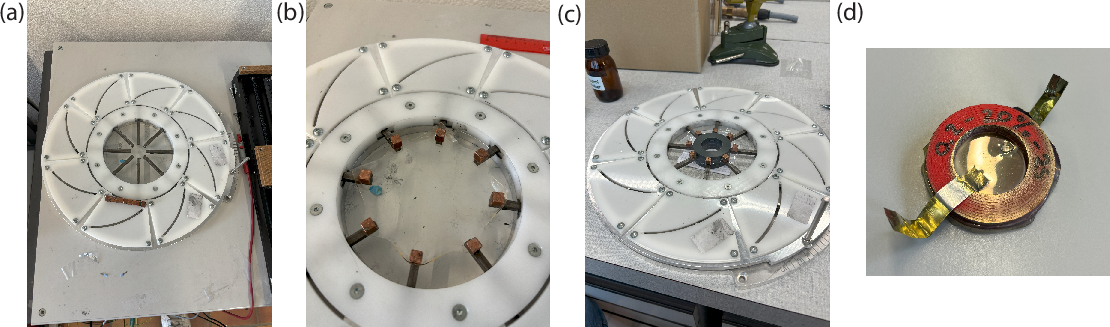}
     \caption{Main phases of the elastomer sample preparation. (a) placement of the unstretched membrane in the mechanical diaphragm. (b) Stretched state of the membrane. (c) Placement and gluing of the rings to clamp the membrane in its stretched state. (d) Final sample with sputtered gold electrodes and aluminium foil electrodes for circuit connection.}
     \label{fig:Sample_Prep}
 \end{figure*}

The preparation of the samples includes the pre-stretching of the membrane, its fixation on an external frame, the application of thin electrodes, and the preparation of connections to the voltage source. The workflow for the preparation is well established and is carried out with tools and materials developed in the laboratory for the purpose. The photographs in figure \ref{fig:Sample_Prep} show the main steps in the preparation of a transducer: 1) The silicone membrane is positioned in the iris device and adhered to the eight movable tabs. 2) The movable tabs are shifted to strain the membrane radially (here shown at the maximum radial displacement). 3)  A rigid frame is positioned in contact with the membrane to take up radial stress. 4) After the membrane is fixated on the frame with an adhesive, thin ($\approx 20nm$) gold electrodes are sputtered. Aluminium foil leads were then applied and connected to the two faces. 

\subsection{Test Setup}
\begin{figure}
     \centering
     \includegraphics[width=\columnwidth]{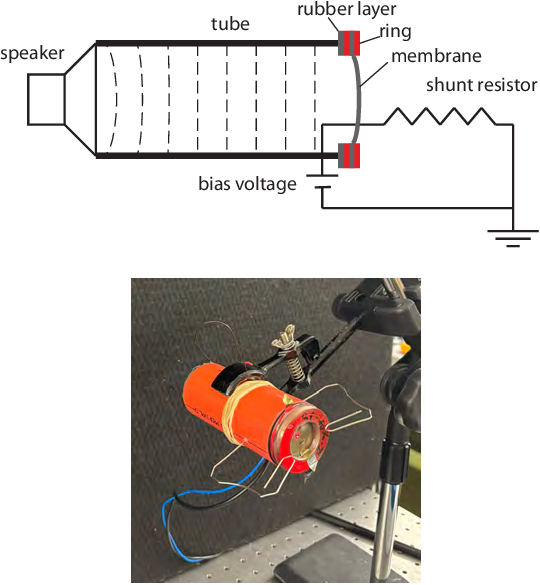}
     \caption{Schematic representation of the test setup (top) and image showing the mounted membrane, where the red vibrometer dot is visible (bottom)}
     \label{fig:Setup}
 \end{figure}

In the experiments, an acoustic wave was shone onto the membrane from a speaker attached to a 200~mm long polymer tube, as schematically shown in the top of Fig.~\ref{fig:Setup}.  The speaker was driven by a Thomann t.amp E400, connected to the analog output of a NI PXI signal generator card. The membrane under test was positioned at the opposite end of the pipe. A laser Doppler vibrometer (LDV, Polytec PDV-100) was used to monitor the velocity of the center of the membrane. A high voltage power supply (Stanford Research Systems PS350) was connected to the leads to the membrane to apply the necessary bias voltage $V_0$. A $1$~M$\Omega$ shunt resistor $R_s$ between the outer electrode and ground was used to measure the capacitance dependent current flowing to the biased membrane.

\subsection{Measurements}
Tests were performed with a sine sweep signal of 5~s duration spanning the frequency range 0-1200~Hz. Within this range, the frequency response of the speaker and the amplifier were assumed to be linear. The membrane velocity, recorded with the LDV, the shunt voltage, and the voltage signal driving the speaker $V_S$ were measured with a rate of 20~kS/s. Tests were at various bias voltage levels ranging from 0~V to 1000~V for membranes with two different levels of pre-stretch, $\tau_0=0.1$ and $\tau_0=0.2$.

\subsection{Results and discussion}
\begin{figure}
     \centering
     \includegraphics[width=\columnwidth]{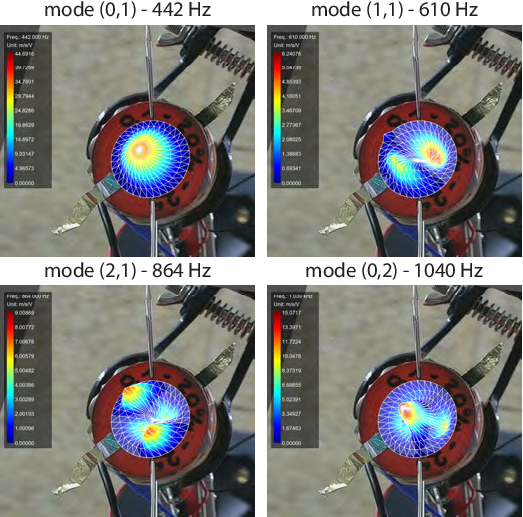}\\
     \caption{First 4 mode shapes for a 20~$\mu$m membrane submitted to 20\% static strain, measured by an SLDV}
     \label{fig:Scanning_vib}
 \end{figure}

In the measured frequency range, four membrane mode shapes can be discerned, as shown in Fig.~\ref{fig:Scanning_vib}. The shapes are recorded with an Optoscan laser scanning vibrometer, and the first (0,1) mode is clearly similar to the modelled shape shown in Fig.~\ref{fig:Membrane_Model}. The (0,2) mode reveals that the mode shape is not perfectly axially symmetric, which hints that the initial strain is not equal along the circumference.

\begin{figure}
     \centering
     \includegraphics[width=\columnwidth]{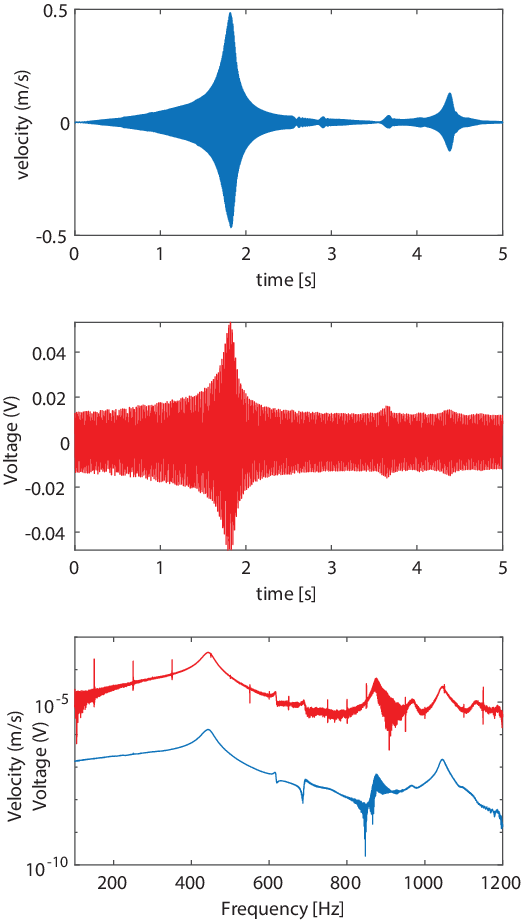}\\
     \caption{Measured time signals (membrane velocity and generated voltage) for a 100~$\mu$m membrane with a 20\% static strain. The lower panel shows their amplitude spectra, where the velocity is recalculated to displacement for comparison with Fig.~\ref{fig:timesig}.}
     \label{fig:overview}
 \end{figure}

Fig.~\ref{fig:overview} shows an overview of the quantities measured for the assessment of the electromechanical coupling capability of the biased membranes. From top to bottom: velocity measured with a single point vibrometer at the center of the membrane, voltage measured over the shunt resistor, and spectra of both time signals. The voltage spectrum shows no frequency-upconversion, the main contribution comes from the first membrane mode. The other membrane modes are equally exciting a voltage although according to Fig.~\ref{fig:timesig} this should not be the case. Both effects can be attributed to the absence of true symmetry in the experimental setup: initial bulging and membrane tension which is not axisymmetric will lead to electro-mechanical coupling for all modes.

\begin{figure*}
     \centering
     \includegraphics[width = \textwidth]{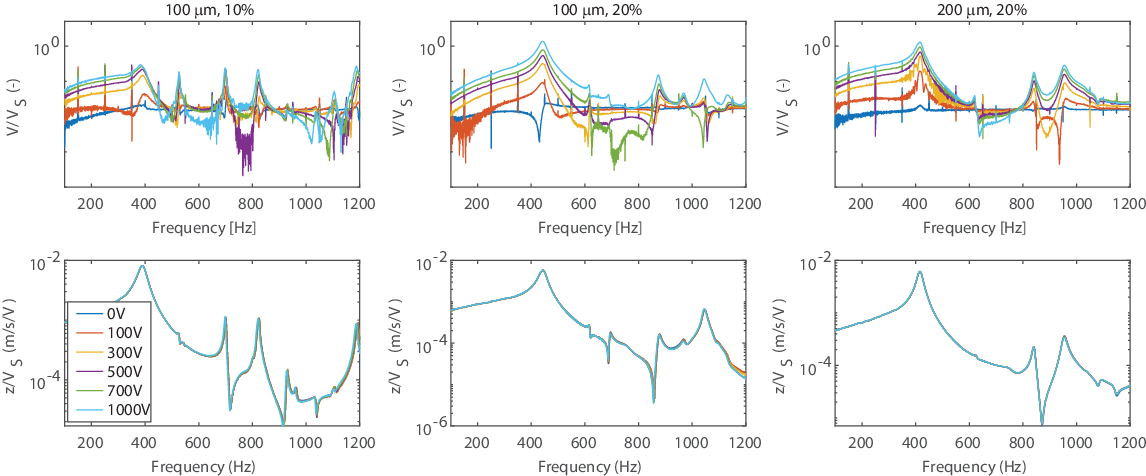}
     \caption{Overview of the measured voltage and displacement transfer functions, normalized by the loudspeaker's input voltage $V_S$. The results are shown for two different membrane thicknesses, and two static strain values of the thinner membrane. The legend shows the range of bias voltage values $V_0$.}
     \label{fig:expresult}
\end{figure*}

From these measurements, it is possible to gain some insights into the coupling of acoustic into electrical power in the systems under consideration. Here, we consider membranes prepared with a pre-stretch $\tau_0=0.1$ and $\tau_0=0.2$ and thickness $100~\mu$m and $200~\mu$m. In all cases, the bias voltage was varied between 0~V and 1000~V, as indicated in the legend of Fig.~\ref{fig:expresult}. 

The bottom panels in Fig.~\ref{fig:expresult} shows the displacement/driving voltage $V_S$ (as a proxy for the acoustic pressure) transfer function. This indicates a clear (0,1)-mode resonance at approximately 390~Hz for the $100 \mu m$ with 10\% prestrain, and 443~Hz and 415~Hz respectively for the membranes with $100 \mu m$ and $200 \mu m$ thickness stretched to 20\%. The two membranes stretched to 20\% do not have the same eigenfrequency as predicted numerically, probably because there was some slip during the stretching process and the thicker membrane was therefore not fully stretched. The displacement transfer functions for each membrane are essentially identical for all bias voltage values. This indicates that the electric field resulting from the potential applied across the membranes is not sufficient to deform them in an appreciable manner, and consequently they do not affect the membrane tension after the pre-stretch. Assuming a nearly incompressible behavior of the silicone, a radial stretch of 20\% leads to a thickness reduction of 30\%. Accordingly, the field across the $100~\mu$m membrane will be ranging from $0$~V/m to $14.4$~MV/m, fairly modest values as compared to the fields of the order of $100$~MV/m \cite{dong2016application} applied to modify the resonant frequency of similar membranes. 
 
The shunt voltage/driving voltage transfer functions presented in the top panels show a clear sensitivity to the applied bias voltage. This observation is compatible with the relation described in (\ref{eq:current}) and the numerical simulations. The two membranes stretched to 20\% clearly generated higher voltages than the membrane stretched to 10\% although the dynamic displacement $z$ is consistently larger for the latter membrane. Since the numerical models do not predict significant voltage amplitude differences with increasing strain, the reason must be a larger initial out-of-plane deformation. Given the fact that the membranes are stretched in 8 points along the circumference, rather than over the entire edge, it can be expected that local buckling occurs. However, this could not be investigated with the measurements at hand.
 
 \section{Conclusions}
The simulations and experiments described in this work show that the combination of thin dielectrics with an externally applied bias voltage show coupling between the mechanical and electrical domain, similar to a piezoelectric material. However, the mechanism at the base of the observed flow of current is different than in piezoelectric crystals with spontaneous electric dipoles. In the membranes we investigated, the electric field acting on them was approximately one order of magnitude smaller than the fields typically reported for actuation applications. Hence, the observed coupling did not affect the mechanical response of the system. 

In both possible applications outlined in the introduction (meta-surface and stack actuator), the ability to functionalize a soft dielectric material to achieve electromechanical coupling offers new possibilities, where stiffer and more brittle materials such as ceramics (or dielectric polymers such as PVDF) do not fulfil requirements such as the need to undergo large strains or to match the mechanical impedance to a specific application (such as in acoustics). Additionally, the experiments discussed here show the ability to tune the electromechanical coupling coefficient of the elements by varying the level of applied bias voltage. This additional degree of freedom can be used to adapt the response of smart systems. 

\bibliographystyle{unsrt}
\bibliography{References}

@article{cheng2023recent,
  title={Recent advances of capacitive sensors: Materials, microstructure designs, applications, and opportunities},
  author={Cheng, Allen J and Wu, Liao and Sha, Zhao and Chang, Wenkai and Chu, Dewei and Wang, Chun Hui and Peng, Shuhua},
  journal={Advanced Materials Technologies},
  volume={8},
  number={11},
  pages={2201959},
  year={2023},
  publisher={Wiley Online Library}
}

@article{qin2021flexible,
  title={Flexible and stretchable capacitive sensors with different microstructures},
  author={Qin, Jing and Yin, Li-Juan and Hao, Ya-Nan and Zhong, Shao-Long and Zhang, Dong-Li and Bi, Ke and Zhang, Yong-Xin and Zhao, Yu and Dang, Zhi-Min},
  journal={Advanced Materials},
  volume={33},
  number={34},
  pages={2008267},
  year={2021},
  publisher={Wiley Online Library}
}

@article{habib2022review,
  title={A review of ceramic, polymer and composite piezoelectric materials},
  author={Habib, Mahpara and Lantgios, Iza and Hornbostel, Katherine},
  journal={Journal of Physics D: Applied Physics},
  volume={55},
  number={42},
  pages={423002},
  year={2022},
  publisher={IOP Publishing}
}

@article{sekhar2023review,
  title={A review on piezoelectric materials and their applications},
  author={Sekhar, Madunuri Chandra and Veena, Eshwarappa and Kumar, Nagasamudram Suresh and Naidu, Kadiyala Chandra Babu and Mallikarjuna, Allam and Basha, Dudekula Baba},
  journal={Crystal Research and Technology},
  volume={58},
  number={2},
  pages={2200130},
  year={2023},
  publisher={Wiley Online Library}
}

@article{safaei2019review,
  title={A review of energy harvesting using piezoelectric materials: state-of-the-art a decade later (2008--2018)},
  author={Safaei, Mohsen and Sodano, Henry A and Anton, Steven R},
  journal={Smart materials and structures},
  volume={28},
  number={11},
  pages={113001},
  year={2019},
  publisher={IOP publishing}
}

@article{khoo20153d,
  title={3D printing of smart materials: A review on recent progresses in 4D printing},
  author={Khoo, Zhong Xun and Teoh, Joanne Ee Mei and Liu, Yong and Chua, Chee Kai and Yang, Shoufeng and An, Jia and Leong, Kah Fai and Yeong, Wai Yee},
  journal={Virtual and Physical Prototyping},
  volume={10},
  number={3},
  pages={103--122},
  year={2015},
  publisher={Taylor \& Francis}
}

@article{bahl2020smart,
  title={Smart materials types, properties and applications: A review},
  author={Bahl, Shashi and Nagar, Himanshu and Singh, Inderpreet and Sehgal, Shankar},
  journal={Materials Today: Proceedings},
  volume={28},
  pages={1302--1306},
  year={2020},
  publisher={Elsevier}
}

@article{vonSzczepanski2023high,
  title={High-Permittivity Polysiloxanes for Solvent-Free Fabrication of Dielectric Elastomer Actuators},
  author={von Szczepanski, Johannes and Opris, Dorina M},
  journal={Advanced Materials Technologies},
  volume={8},
  number={2},
  pages={2201372},
  year={2023},
  publisher={Wiley Online Library}
}

@inproceedings{munz1998deformation,
  title={Deformation and strength behavior of a soft PZT ceramic},
  author={Munz, Dietrich and Fett, Theo and Mueller, Stefan and Thun, Gerhard},
  booktitle={Smart Structures and Materials 1998: Mathematics and Control in Smart Structures},
  volume={3323},
  pages={84--95},
  year={1998},
  organization={SPIE}
}

@article{van2024implementation,
  title={Implementation of tunable frequency-dependent stiffness elements via integrated shunted piezoelectric stacks},
author={Van Damme, Bart and  Weber, Rico and Schmied, Jascha U and Spierings, Adriaan and  Bergamini, Andrea},
  journal={Smart Materials and Structures},
  volume={33},
  number={7},
  pages={075037},
  year={2024},
  publisher={IOP Publishing}
}

@article{bouzid2005pzt,
  title={PZT phase diagram determination by measurement of elastic moduli},
  author={Bouzid, Abderrazak and Bourim, EM and Gabbay, Maurice and Fantozzi, Gilbert},
  journal={Journal of the European Ceramic Society},
  volume={25},
  number={13},
  pages={3213--3221},
  year={2005},
  publisher={Elsevier}
}

@article{dmitriev2006dependence,
  title={Dependence of the dielectric constant on the structure of extruded polyvinylidene fluoride films},
  author={Dmitriev, I Yu and Gladchenko, SV and Lavrent’ev, VK and Praslova, OE and Elyashevich, GK},
  journal={Russian journal of applied chemistry},
  volume={79},
  pages={642--646},
  year={2006},
  publisher={Springer}
}

@article{owusu2023make,
  title={How to make elastomers piezoelectric?},
  author={Owusu, Francis and Venkatesan, Thulasinath Raman and N{\"u}esch, Frank A and Negri, Ricardo Martin and Opris, Dorina M},
  journal={Advanced Materials Technologies},
  volume={8},
  number={15},
  pages={2300099},
  year={2023},
  publisher={Wiley Online Library}
}

@article{kovacs2009stacked,
  title={Stacked dielectric elastomer actuator for tensile force transmission},
  author={Kovacs, Gabor and D{\"u}ring, L and Michel, S and Terrasi, G},
  journal={Sensors and actuators A: Physical},
  volume={155},
  number={2},
  pages={299--307},
  year={2009},
  publisher={Elsevier}
}

@inproceedings{dong2016application,
  title={Application of bias voltage to tune the resonant frequency of membrane-based electroactive polymer energy harvesters},
  author={Dong, Lin and Grissom, Michael and Fisher, Frank T},
  booktitle={Energy Harvesting and Storage: Materials, Devices, and Applications VII},
  volume={9865},
  pages={96--110},
  year={2016},
  organization={SPIE}
}

@article{Kim2019,
   author = {Joshua Kim and En Fan Chou and Jamie Le and Sabrina Wong and Michael Chu and Michelle Khine},
   doi = {10.1002/adhm.201900109},
   issn = {21922659},
   issue = {13},
   journal = {Advanced Healthcare Materials},
   month = {7},
   pmid = {31033256},
   publisher = {Wiley-VCH Verlag},
   title = {Soft Wearable Pressure Sensors for Beat-to-Beat Blood Pressure Monitoring},
   volume = {8},
   year = {2019}
}

@misc{Chen2020,
   author = {Wufan Chen and Xin Yan},
   doi = {10.1016/j.jmst.2019.11.010},
   issn = {10050302},
   journal = {Journal of Materials Science and Technology},
   month = {4},
   pages = {175-188},
   publisher = {Chinese Society of Metals},
   title = {Progress in achieving high-performance piezoresistive and capacitive flexible pressure sensors: A review},
   volume = {43},
   year = {2020}
}

@article{Zhao2023,
   author = {Jiawei Zhao and Haoyu Guo and Haiyang Liu and Tongqiang Fu and Wenzhe Zhou and Zicai Zhu and Qiao Hu},
   doi = {10.1021/acsami.3c10100},
   issn = {19448252},
   issue = {40},
   journal = {ACS Applied Materials and Interfaces},
   month = {10},
   pages = {47327-47337},
   publisher = {American Chemical Society},
   title = {Carbon Nanotube Network Topology-Enhanced Iontronic Capacitive Pressure Sensor with High Linearity and Ultrahigh Sensitivity},
   volume = {15},
   year = {2023}
}

@misc{Ha2022,
   author = {Kyoung Ho Ha and Heeyong Huh and Zhengjie Li and Nanshu Lu},
   doi = {10.1021/acsnano.2c00308},
   issn = {1936086X},
   issue = {3},
   journal = {ACS Nano},
   month = {3},
   pages = {3442-3448},
   pmid = {35262335},
   publisher = {American Chemical Society},
   title = {Soft Capacitive Pressure Sensors: Trends, Challenges, and Perspectives},
   volume = {16},
   year = {2022}
}

@article{Zammali2022,
   author = {Marouen Zammali and Sijun Liu and Wei Yu},
   doi = {10.1002/admi.202200015},
   issn = {21967350},
   issue = {17},
   journal = {Advanced Materials Interfaces},
   month = {6},
   publisher = {John Wiley and Sons Inc},
   title = {A Flexible, Transparent, Ultralow Detection Limit Capacitive Pressure Sensor},
   volume = {9},
   year = {2022}
}

@article{Kumaresan2022,
   author = {Yogeenth Kumaresan and Sihang Ma and Oliver Ozioko and Ravinder Dahiya},
   doi = {10.1109/JSEN.2022.3143030},
   issn = {15581748},
   issue = {5},
   journal = {IEEE Sensors Journal},
   month = {3},
   pages = {3974-3982},
   publisher = {Institute of Electrical and Electronics Engineers Inc.},
   title = {Soft Capacitive Pressure Sensor with Enhanced Sensitivity Assisted by ZnO NW Interlayers and Airgap},
   volume = {22},
   year = {2022}
}

@article{Dagdeviren2014,
   author = {Canan Dagdeviren and Yewang Su and Pauline Joe and Raissa Yona and Yuhao Liu and Yun Soung Kim and Yongan Huang and Anoop R. Damadoran and Jing Xia and Lane W. Martin and Yonggang Huang and John A. Rogers},
   doi = {10.1038/ncomms5496},
   issn = {20411723},
   journal = {Nature Communications},
   month = {8},
   pmid = {25092496},
   publisher = {Nature Publishing Group},
   title = {Conformable amplified lead zirconate titanate sensors with enhanced piezoelectric response for cutaneous pressure monitoring},
   volume = {5},
   year = {2014}
}

@article{Owusu2022,
   author = {Francis Owusu and Frank A. Nüesch and Dorina M. Opris},
   doi = {10.1002/adfm.202207083},
   issn = {16163028},
   issue = {41},
   journal = {Advanced Functional Materials},
   month = {10},
   publisher = {John Wiley and Sons Inc},
   title = {Stretchable High Response Piezoelectric Elastomers Based on Polable Polynorbornene Fillers in a Polydimethylsiloxane Matrix},
   volume = {32},
   year = {2022}
}

@article{jin2021skin,
  title={Skin-like elastomer embedded zinc oxide nanoarrays for biomechanical energy harvesting},
  author={Jin, Congran and Dong, Lin and Xu, Zhe and Closson, Andrew and Cabe, Andrew and Gruslova, Aleksandra and Jenney, Scott and Escobedo, Danny and Elliott, James and Zhang, Michael and others},
  journal={Advanced Materials Interfaces},
  volume={8},
  number={10},
  pages={2100094},
  year={2021},
  publisher={Wiley Online Library}
}

@misc{Kumar2024,
   author = {Ashish Kumar and Arathy Varghese and Dheeraj Kalra and Anshuman Raunak and Jaiverdhan and Mahanth Prasad and Vijay Janyani and R. P. Yadav},
   doi = {10.1016/j.mssp.2023.107879},
   issn = {13698001},
   journal = {Materials Science in Semiconductor Processing},
   month = {1},
   publisher = {Elsevier Ltd},
   title = {MEMS-based piezoresistive and capacitive microphones: A review on materials and methods},
   volume = {169},
   year = {2024}
}

@article{Shin2017,
   author = {Sung Ho Shin and Sangyoon Ji and Seiho Choi and Kyoung Hee Pyo and Byeong Wan An and Jihun Park and Joohee Kim and Ju Young Kim and Ki Suk Lee and Soon Yong Kwon and Jaeyeong Heo and Byong Guk Park and Jang Ung Park},
   doi = {10.1038/ncomms14950},
   issn = {20411723},
   journal = {Nature Communications},
   month = {3},
   pmid = {28361867},
   publisher = {Nature Publishing Group},
   title = {Integrated arrays of air-dielectric graphene transistors as transparent active-matrix pressure sensors for wide pressure ranges},
   volume = {8},
   year = {2017}
}
\end{document}